\documentclass[10pt,conference]{IEEEtran}
\newcommand{\gmha}{\gamma_{1}^{h}}
\newcommand{\gmhb}{\gamma_{2}^{h}}
\newcommand{\gmga}{\gamma_{1}^{g}}
\newcommand{\gmgb}{\gamma_{2}^{g}}
\newcommand{\phph}{\phi_{P_{1},P_{2}}^{h}}
\newcommand{\phqg}{\phi_{Q_{1},Q_{2}}^{g}}
\newcommand{\phpg}{\phi_{P_{1},P_{2}}^{g}}
\newcommand{\phqh}{\phi_{Q_{1},Q_{2}}^{h}}
\newtheorem{thm}{Theorem}[section]
\usepackage{epsfig}
\usepackage{graphicx}
\usepackage{epstopdf}
\usepackage{amsmath}
\usepackage{amsfonts}
\usepackage{amssymb}

\begin{document}

\title{Achievable Secrecy Sum-Rate in a Fading MAC-WT with Power Control and without CSI of Eavesdropper}
\author{
\IEEEauthorblockN{Shahid M Shah, Vireshwar Kumar and Vinod Sharma}
\IEEEauthorblockA{Department of Electrical Communication Engineering\\
Indian Institute of Science, 
Bangalore 560012, India\\
Email: shahid@ece.iisc.ernet.in, vireshwar@pal.ece.iisc.ernet.in, vinod@ece.iisc.ernet.in}
}
\maketitle

\begin{abstract}
We consider a two user fading Multiple Access Channel with a wire-tapper (MAC-WT) where the transmitter has the channel state information (CSI) to the intended receiver but not to the eavesdropper (eve). We provide an achievable secrecy sum-rate with power control. We next provide a secrecy sum-rate with power control and cooperative jamming (CJ). We also study an achievable secrecy sum rate by employing an ON/OFF power control scheme which is more easily computable. Finally we employ CJ over this power control scheme. Results show that CJ boosts the secrecy sum-rate significantly even if we do not know the CSI of the eve's channel. At high SNR, the secrecy sum-rate (with CJ) without CSI of the eve exceeds the secrecy sum-rate (without CJ) with full CSI of the eve.
\end{abstract}


\section{Introduction}
Security is one of the most important considerations in transmission of information from one user to another. It involves confidentiality, integrity, authentication and non-repudiation \cite{liang}. We will be concerned about confidentiality. This guarantees that the legitimate users successfully receive the information intended for them while any eavesdropper is not able to interpret this information. We will be concerned with the eavesdroppers who are passive attackers, e.g., they attempt to interpret the transmitted information without injecting any new information or trying to modify the information transmitted.

Traditional techniques to achieve confidentiality in this setup are based on cryptographic encryption (\cite{menezes}, \cite{stallings}). However now, Information Theoretical Security is also being actively studied (\cite{liang}, \cite{trappe}). This does not require the secret/public keys used in cryptographic techniques. Key management, especially for wireless channels can be very challenging. Also, information theoretic security unlike the cryptography based techniques can provide provably secure communication. Information theoretic security can also be used in a system in addition to cryptographic techniques to add additional layers of protection to the information transmission or to achieve key agreement and/or distribution.

The Information theoretic approach for secrecy systems was first investigated by Shannon \cite{shannon1949} in 1949. Wyner \cite{wyner1975} considers communicating a secret over a wiretap channel in the form of degraded broadcast channels, without using a key. Wyner's work was in turn extended by Leung and Hellman \cite{hellman1976} to the Gaussian channel.   Csisz\`{a}r and K\"{o}rner \cite{csizar1980} considers a general discrete memoryless broadcast channel, and shows that the secrecy capacity is positive if the main channel to the intended user is more capable than of the eavesdropper, and zero if the wiretapper's channel is less noisy. Secrecy capacity of MIMO (Multiple Input Multiple Output) channels was obtained in \cite{khisti}, \cite{babak2008}  and \cite{shamai2009}. In \cite{negi2007}, an artificial noise concept is proposed for MIMO channels to enhance the secrecy rate even when the eavesdropper's channel is better than the main channel. In \cite{he2011}, this result has been extended to the case where no CSI of the eavesdropper is available. The fading channel is studied in \cite{gopala2008} where power allocation schemes without CSI of the eavesdropper's channel to the transmitter are also obtained. In \cite{bloch2008}, a wire-tap channel with slow fading is studied where an outage analysis with full CSI of the eavesdropper and imperfect CSI of the eavesdropper is performed. Practical codes for the single user wire-tap channel were first reported in \cite{thangaraj}. Liang et al.\cite{broadcast} have studied a broadcast channel with a wire-tapper. Interference channels with confidential messages are studied in \cite{interference}.

 The first results in information theoretic security for a Multiple Access Channel (MAC) were obtained in \cite{yingbin2008} and \cite{yener2006}. In \cite{yingbin2008}, each user treats the other as an eavesdropper while in \cite{yener2006}, the eavesdropper is at the receiving end. In \cite{cooperative}, Yener and Tekin propose a technique called cooperative jamming in which a user that is not transmitting, can send a jamming signal so that the eavesdropper is more confused. This significantly improves the secrecy rate region. A fading MAC was also studied by Yener and Tekin \cite{yener2007}, where they assume that the CSI of the eavesdropper's channel is perfectly known at the transmitting users. In \cite{ulukus2010}, Bassily and Ulukus have proposed Ergodic Secret Alignment to further improve the secrecy sum-rate region of the MAC with an eavesdropper.

In this paper, we consider a fading MAC-WT assuming no CSI of the eavesdropper at the transmitting users. Since the eavesdropper may not transmit any signal (it is passive), the transmitters often will not know its channel. We obtain a power control scheme that maximizes the sum secrecy rate and then also employ cooperative jamming over this scheme. It will be shown that cooperative jamming can significantly increase the secrecy rate. But these policies are difficult to compute. Thus, next we consider a computationally simpler ON/OFF power control policy. We obtain its thresholds to maximize the secrecy sum-rate. Finally, we also incorporate cooperative jamming over this power control policy. With this, at high SNR, the secrecy sum-rate exceeds the sum-rate when CSI of the eavesdropper is perfectly known at the transmitter but the cooperative jamming is not used.

The rest of the paper is organized as follows: In Section II, we define the channel model and state the problem. In Section III, we obtain the power control policy with and without cooperative jamming. Section IV discusses ON/OFF power control policy with and without cooperative jamming. In Section V, we compare the different policies numerically. Finally in Section VI, we conclude this paper and discuss the future work. 

\section{Channel Model And Problem Statement}
We consider a system with two users who want to communicate over a fading MAC to a legitimate receiver. There is also an eavesdropper who is trying to get access to the output received by the legitimate receiver. Transmitter $k=1,2$ chooses message $W_{k}$ for transmission from a set $\mathcal{W}_{k} = \{1, 2,..., M_{k}\}$ with uniform distribution. These messages are encoded into $\{X_{k,1}, ..., X_{k,n}\}$ using $(2^{nR_{k}}, n)$ codes. The legitimate receiver gets $Y_{i}$ and the eavesdropper gets $Z_{i}$ at time $i$. The decoder at the legitimate receiver estimates the transmitted  message as $\tilde{W} = (\tilde{W_{1}}, \tilde{W_{2}})$ from $\textbf{Y}^{n} \equiv \{Y_{1}, ..., Y_{n}\}$. The legitimate receiver should receive the message reliably while the eavesdropper should not be able to decode it. It is assumed that the legitimate receiver as well as the eavesdropper know the codebooks.

The channel model can be mathematically represented as:
\begin{equation}
\label{model1}
Y_{i}= \tilde{h}_{1,i}X_{1,i} + \tilde{h}_{2,i}X_{2,i} + N_{R,i}
\end{equation}
\begin{equation}
\label{model2}
Z_{i}=\tilde{g}_{1,i}X_{1,i} + \tilde{g}_{2,i}X_{2,i} +  N_{E,i}
\\
\end{equation}
where $\tilde{h}_{k,i}, ~\tilde{g}_{k,i}$ are the complex channel gains from the transmitter $k$ to the legitimate receiver and the eavesdropper respectively. Also $\{N_{R,i}\}$ and $\{N_{E,i}\}$ are Complex additive white Gaussian noise (AWGN) each with circularly symmetric independent components distributed as $\mathcal{N}(0,1)$, where $\mathcal{N}(a,b)$ is Gaussian distribution with mean $a$ and variance $b$. Also we define $\vert\tilde{h}_{k,i}\vert^{2} = h_{k,i}$ and $\vert\tilde{g}_{k,i}\vert^{2} = g_{k,i}$, for $k=1, 2$.
We assume that $\{h_{k,i}, i\geq 1\}$ and $\{g_{k,i}, i\geq 1\}$ are independent, identically distributed (iid), and that each sequence is independent of the other (thus the channels experience flat, fast fading).
We also assume the power constraints:
\begin{equation}
\label{p_cons}
\frac{1}{n} \sum\limits_{i=1}^{n} X^{2}_{ki} \leq \bar{P}_{k}, ~k=1,2.
\end{equation}

The equivocation rate used in this paper is as defined in \cite{yener2006}. We use collective secrecy constraint to take the multi-access nature of the channel into account. Define\\
\begin{equation}
\Delta _{L}^{n} = \frac{H(W_{L}\vert Z^{n})}{H(W_{L})} 
\end{equation}
where  $L \subseteq \{1, 2\}$, $Z^{n} = (Z_{1}, ... ,Z_{n})$. For each $n$ we need codebooks such that the average probability of error to the legitimate receiver goes to zero and $\Delta _{L}^{n} \rightarrow 1$ as $n \rightarrow \infty$ for each $L  ~\subseteq \{1,2\}$.
Also let $h=(h_{1}, h_{2}), g=(g_{1}, g_{2})$, where $h_i = \vert\tilde{h}_{i}\vert^{2}$, $g_i = \vert\tilde{g}_{i}\vert^{2}, i=1,2$ are exponentially distributed.

Then from \cite{yener2006}, if the CSI $h_{k}, g_{k}$ are known at both the transmitters at time $i$, the secrecy rate region for this case is:
\begin{equation}
\label{r1_fullcsi}
R_{1} \leq \mathsf{E}_{h,g}\left\lbrace\left[log\frac{(1+h_{1}P_{1}(h,g))(1+g_{2}P_{2}(h,g))}{1+g_{1}P_{1}(h,g)+g_{2}P_{2}(h,g)}\right]^{+}\right\rbrace,
\end{equation}

\begin{equation}
\label{r2_fullcsi}
R_{2}\leq \mathsf{E}_{h,g}\left\lbrace\left[log\frac{(1+g_{1}P_{1}(h,g))(1+h_{2}P_{2}(h,g))}{1+g_{1}P_{1}(h,g)+g_{2}P_{2}(h,g)}\right]^{+}\right\rbrace,
\end{equation}

\begin{equation}
\label{rs_fullcsi}
R_{1}+R_{2} \leq \mathsf{E}_{h,g}\left\lbrace\left[log\frac{1+h_{1}P_{1}(h,g)+h_{2}P_{2}(h,g)}{1+g_{1}P_{1}(h,g)+g_{2}P_{2}(h,g)}\right]^{+}\right\rbrace,
\end{equation}
where $P_{1}(h,g)$ and $P_{2}(h,g)$ are the transmit powers satisfying the constraint (3)and Gaussian signalling is used.

In \cite{yener2006}, the optimal power allocation policy which maximizes the sum secrecy rate (7) has been found. In this paper, we extend this result to the case when the CSI of the legitimate receiver is known but the CSI of the eavesdropper may not be known at the transmitter; only the distribution is known. Since we are assuming a passive eavesdropper, this will often be a more reasonable assumption, i.e., there is no transmission from the eavesdropper to the transmitters for them to estimate its channel.

\section{Power Control with main CSI only}
\subsection{Power control without Cooperative Jamming}
In this section we consider power control which maximizes the sum secrecy rate when only the main channel (to the legitimate user) CSI is known at the transmitters. Let $P_{k}(h)$ be the power used by a policy when the main channel gain is $h=(h_1, h_2)$. Of course, the policy should satisfy the average power constraint (\ref{p_cons}).
We need the following notation
\begin{equation}
\label{notation}
\phi_{p_{1},p_{2}}^{s} = 1 + s_{1}p_{1} + s_{2}p_{2},
\end{equation}
where $s$ is the channel state ($h$ or $g$) and $p_{k}$ is the power used.
The following theorem can be proved as in \cite{gopala2008}.
\\
\begin{thm}
For a given power control policy $\{P_{k}(h)\}, ~k=1, 2$, the following secrecy sum-rate
\begin{equation}
\label{rs_nocsi}
\mathsf{E}_{h,g}\left\lbrace\left[log\left(\frac{\phph}{\phpg}\right)\right]^{+}\right\rbrace
\end{equation}
is achievable.
\end{thm}

The policy that maximizes (\ref{rs_nocsi}) is not available in closed form, but can be numerically computed (see Appendix). An example will be provided in Section VI. We will also consider a simpler ON/OFF power control policy which was employed in \cite{gopala2008} for a single user case.

Next we consider power control with cooperative jamming.
\subsection{Power Control with Cooperative Jamming}
The power policy obtained in the last section depends on $h=(h_1,h_2)$. If both the main channels $h_{1}$ and $h_{2}$ are good, both the transmitters send their coded symbols. If a transmitter's channel is bad, it may not. In \cite{cooperative} and \cite{yener2007}, it is suggested that when a transmitter is not sending its data, it can help the other user by jamming the channel to the eavesdropper. We extend their result to our set-up.

Let $\{P_{k}(h)\},~ k=1, 2$, be the power control policy when the users are transmitting and  $\{Q_{k}(h)\},~ k=1, 2$, be the power control policy when the users are jamming. To satisfy (3), we need
\begin{equation}
\label{power_cons}
\mathsf{E}_{h}[P_{k}(h) + Q_{k}(h)] \leq \bar{P_{k}},~k=1,2.
\end{equation}
Then we have the following theorem
\\
\begin{thm} With the above power control policies secrecy sum-rate

\setlength{\arraycolsep}{0.0em}
\begin{equation}
\label{rs_coop}
 \mathsf{E}_{h,g}\left\lbrace\left[log\left(\frac{\phph +\phqh -1}{\phpg + \phqg -1}\right)\left(\frac{\phqg}{\phqh}\right)\right]^{+}\right\rbrace
\end{equation}
is achievable.
\end{thm}
The proof of this theorem is given in Appendix A.1

We will obtain the power control policy that maximizes the sum rate in the Appendix. We will see in Section VI that cooperative jamming can significantly improve the sum-rate.

We also propose a simple ON/OFF power control policy with cooperative jamming.
\section{Fading MAC with ON/OFF Power Control}
The policy obtained in Section III can be computed only numerically and its structure is not known. The following ON/OFF policy is easier to compute and is intuitive:

User $k$ transmits with a constant power $P_{k}$ if $h_{k} \geq \tau_{k}$, where $\tau_{k}$ is an appropriate threshold. Hence the following cases arise:
\begin{enumerate}
\item h$_{1}\geq\tau_{1}$, h$_{2}\geq\tau_{2}$ : Both transmit;
\item h$_{1}\geq\tau_{1}$, h$_{2}<\tau_{2}$ : User-1 transmits;
\item h$_{1}<\tau_{1}$, h$_{2}\geq\tau_{2}$ : User-2 transmits;
\item h$_{1}<\tau_{1}$, h$_{2}<\tau_{2}$ : No user transmits.
\end{enumerate}

From average power constraint we get:
\begin{equation}
\label{p_1_on_off}
\bar{P_{1}}=P_{1}Pr(h_{1}\geq\tau_{1})
\end{equation}
and
\begin{equation}
\label{p_2_on_off}
\bar{P_{2}}=P_{2}Pr(h_{2}\geq\tau_{2}).
\end{equation}
Let 
\begin{equation*}
A_{1} \triangleq \{h_{1}\geq\tau_{1},~h_{2}<\tau_{2}\},~
A_{2} \triangleq \{h_{1}<\tau_{1},~h_{2}\geq\tau_{2}\}
\end{equation*}
and
\begin{equation}
\label{indicator}
A_{12} \triangleq \{h_{1}>\tau_{1},~h_{2}>\tau_{2}\}
\end{equation}

The secrecy sum-rate by this policy is given by

\begin{equation*}
R_{B}=\mathsf{E}_{h,g}\left\lbrace\left[log\left(\frac{\phph}{\phpg}\right)1_{A_{12}}\right]^{+}\right\rbrace
\end{equation*}

\begin{equation*}
+~\mathsf{E}_{h,g}\left\lbrace\left[log\left(\frac{\phi_{P_{1},0}^{h}}{\phi_{P_{1},0}^{g}}\right)1_{A_{1}}\right]^{+}\right\rbrace
\end{equation*}
\begin{equation}
\label{r1_onoff}
+~\mathsf{E}_{h,g}\left\lbrace\left[log\left(\frac{\phi_{0,P_{2}}^{h}}{\phi_{0,P_{2}}^{g}}\right)1_{A_{2}}\right]^{+}\right\rbrace		
\end{equation}
where $1_{A}$ is the indicator function of set A.

When $h_{k}$ and $g_{k}$, $k=1,2$ have exponential distributions with independent components, with joint density of $h$ and $g$ respectively given as:
\begin{equation}
f_{1}(h) =\frac{1}{\gmha\gmhb}e^{-\frac{h_{1}}{\gamma_{1}^{h}}}e^{-\frac{h_{2}}{\gamma_{2}^{h}}}~,
~f_{2}(g) = \frac{1}{\gmga\gmgb}e^{-\frac{g_{1}}{\gamma_{1}^{g}}}e^{-\frac{g_{2}}{\gamma_{2}^{g}}}
\end{equation}
then
\begin{equation}
P_{1} = \bar{P_{1}}e^{\frac{\tau _{1}}{\bar{\gamma}_{1}^{h}}}~,~
P_{2} = \bar{P_{2}}e^{\frac{\tau _{2}}{\bar{\gamma}_{2}^{h}}}
\end{equation}

We numerically obtain the secrecy sum-rate for thresholds $\tau_{1}$ and $\tau_{2}$ which maximize sum-rate (\ref{r1_onoff}).
\section{Fading MAC with ON/OFF Power Control and Cooperative Jamming}
Cooperative jamming in section III B has been found to increase the sum-rate substantially (see Fig 1 below). Therefore, we now use it with the ON/OFF policy studied in Section IV. A user when not transmitting its data jams the channel for the eavesdropper. Also it transmits with different powers taking into account the channel gain of the other user. The following cases arise:\\
\begin{enumerate}
\item h$_{1}\geq\tau_{1}$, h$_{2}\geq\tau_{2}$ : Both transmit with power $P_{1a}$, $P_{2a}$;
\item h$_{1}\geq\tau_{1}$, h$_{2}<\tau_{2}$ : User-1 transmits with power $P_{1b}$, user-2 jams with power $Q_{2}$;
\item h$_{1}<\tau_{1}$, h$_{2}\geq\tau_{2}$ : User-2 transmits with power $P_{2b}$, user-1 jams with power $Q_{1}$;
\item h$_{1}<\tau_{1}$, h$_{2}<\tau_{2}$ : None transmits or jams.
\end{enumerate}

The powers and the thresholds in the above scheme are chosen to satisfy the average power constraints.

With this power control scheme, the secrecy sum-rate is given by:

\begin{equation*}
R_{B}^{CJ}=
\mathsf{E}_{h,g}\left\lbrace\left[log\left(\frac{\phi_{P_{1a},P_{2a}}^{h}}{\phi_{P_{1a},P_{2a}}^{g}}\right)1_{A_{12}}\right]^{+}\right\rbrace
\end{equation*}
\begin{equation*}
~~~~~+~\mathsf{E}_{h,g}\left\lbrace\left[log\left(\frac{\phi_{P_{1b},Q_{2}}^{h}}{\phi_{P_{1b},Q_{2}}^{g}}\right)1_{A_{1}}\right]^{+}\right\rbrace
\end{equation*}
\begin{equation}
\label{rs_coop_on_off}
~~~~~+~\mathsf{E}_{h,g}\left\lbrace\left[log\left(\frac{\phi_{Q_{1},P_{2b}}^{h}}{\phi_{Q_{1},P_{2b}}^{g}}\right)1_{A_{2}}\right]^{+}\right\rbrace
\end{equation}

When $\tilde{h}_{k}$ and $\tilde{g}_{k}$ have Rayleigh distribution\\
\begin{eqnarray}
\bar{P_{1}}&{} ~& = P_{1a}e^{-\frac{\tau _{1}}{\bar{\gamma}_{1}^{h}}}e^{-\frac{\tau _{2}}{\bar{\gamma}_{2}^{h}}}+~P_{1b}e^{-\frac{\tau _{1}}{\bar{\gamma}_{1}^{h}}}(1-e^{-\frac{\tau _{2}}{\bar{\gamma}_{2}^{h}}})\nonumber\\ 
&&{+}\: Q_{1}e^{-\frac{\tau _{2}}{\bar{\gamma}_{2}^{h}}}(1-e^{-\frac{\tau _{1}}{\bar{\gamma}_{1}^{h}}}),
\end{eqnarray}
\begin{eqnarray}
\bar{P_{2}}&{}~ & = P_{2a}e^{-\frac{\tau _{1}}{\bar{\gamma}_{1}^{h}}}e^{-\frac{\tau _{2}}{\bar{\gamma}_{2}^{h}}} + P_{2b}e^{-\frac{\tau _{2}}{\bar{\gamma}_{2}^{h}}}(1-e^{-\frac{\tau _{1}}{\bar{\gamma}_{1}^{h}}})\nonumber\\
&&{+}\:Q_{2}e^{-\frac{\tau _{1}}{\bar{\gamma}_{1}^{h}}}(1-e^{-\frac{\tau _{2}}{\bar{\gamma}_{2}^{h}}}).
\end{eqnarray}

\section{Numerical Results}
In this section, we compare the sum rates obtained via the different power control schemes proposed in this paper. The receiver's AWGN noise has variance 1. The fading for each channel is Rayleigh distributed with parameters $\gmha=\gmhb=\gmga=\gmgb=1$. The sum rates are plotted in Fig.1 for different powers $P_{1}=P_{2}$. We observe that cooperative jamming substantially improves the sum-rate (up to 75\%). Of course, for each case knowledge of the eavesdropper's CSI at the transmitter improves the sum-rate. At high SNR, the cooperative jamming can provide sum-rate without CSI higher than the full CSI case without cooperative jamming. Also, optimal ON/OFF power control is sufficient to recover most of the sum rate achievable by a policy (for no eavesdropper's CSI and no jamming, ON/OFF provides sum-rate very close to the optimal algorithm in that scenario). 

\begin{figure}
\epsfig{figure=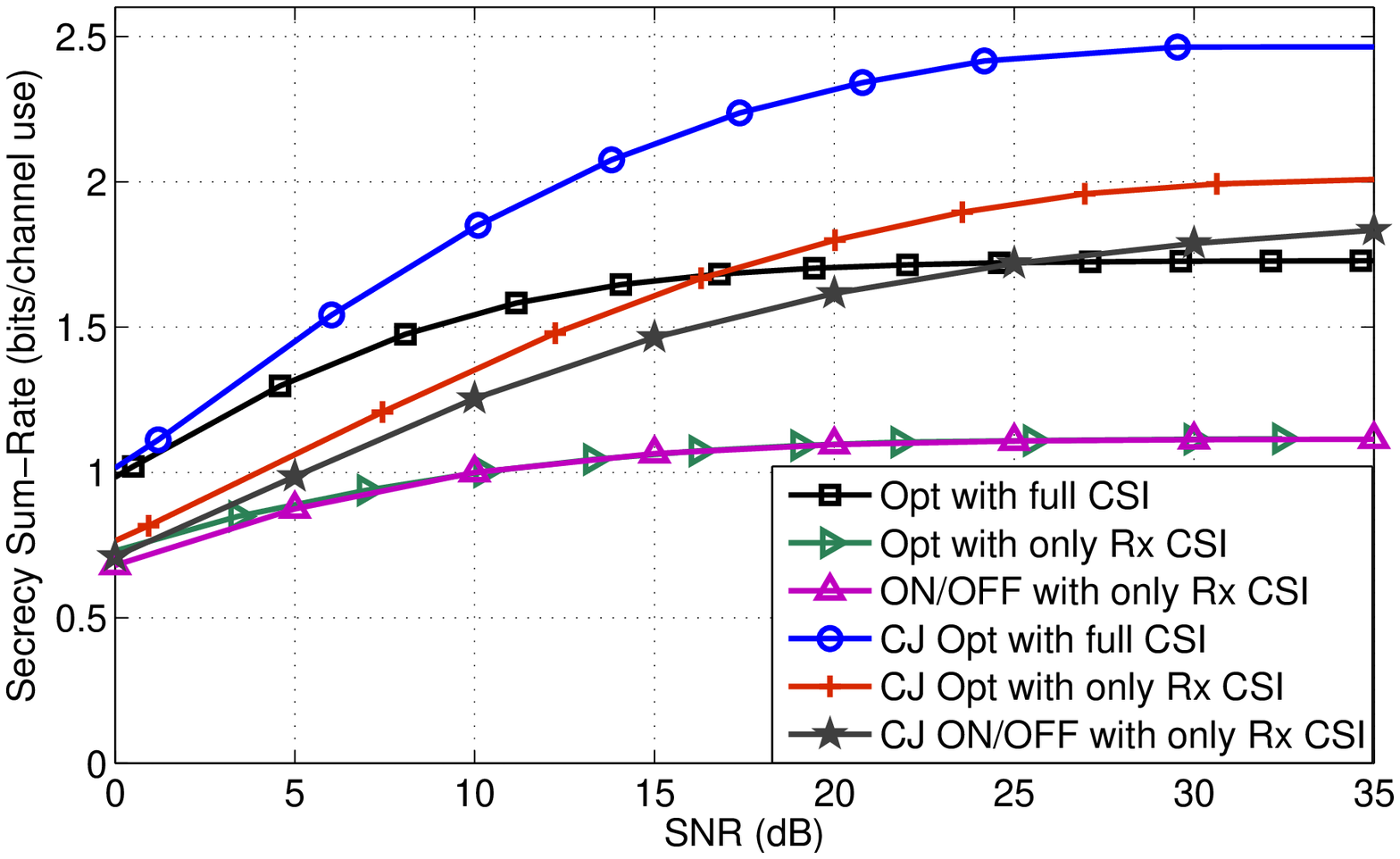,height=6cm,width=16cm}
\caption{Comparison of Full CSI, Only Receiver's CSI  and ON/OFF Power Control policies: With and without Cooperative Jamming}
\label{fig:on_off_opt}
\end{figure}
\section{Conclusions and Future Work}
In this paper, we provide achievable secrecy sum-rate in a fading MAC with an eavesdropper when the eavesdropper's channel is not known to the transmitter. We obtain the power controls that optimize the secrecy sum-rate. We also obtain the power control when cooperative jamming is also employed. It is shown that cooperative jamming can substantially improve the secrecy sum-rates. We, then, obtain more easily computable ON/OFF power control schemes which provide secrecy sum-rates close to the optimal. 

It is shown that via these techniques, one can recover most of the secrecy sum-rate achievable with the perfect knowledge of the CSI of the eavesdropper.

For future work one can consider the schemes when partial CSI of the legitimate receiver's channel is available at the transmitter.

\appendices[Power Control]

\section{Proof of theorems}
\subsection{Proof of theorem 1}
We sketch the outline of proof for sum-rate only, the proof of single user rate follows on the similar lines.
We quantize $h_{1}, h_{2}, g_{1}$ and $g_{2}$ into uniform bins as defined in \cite{gopala2008}, i.e $\left\lbrace h_{1}\right\rbrace _{i_{1}=1}^{q_{1}^{h}}, \left\lbrace h_{2}\right\rbrace _{i_{2}=1}^{q_{2}^{h}}, \left\lbrace g_{1}\right\rbrace _{j_{1}=1}^{q_{1}^{g}} $ and $\left\lbrace g_{2}\right\rbrace _{j_{2}=1}^{q_{2}^{g}} $ where $h_{1}\in [0,M_{1}^{h}], h_{2}\in [0,M_{2}^{h}], g_{1}\in [0,M_{1}^{g}] $ and $ g_{2}\in [0,M_{2}^{g}]$
\\
Also let 
\\
$\mathcal{H}(i_{1},i_{2})$=\{\textbf{h}:$h_{1,i_{1}}\leq~h_{1}\leq~h{1,i_{1}+1},h_{2,i_{2}}\leq~h_{2}\leq h{2,i_{2}+1}$\}
\\
$\mathcal{G}(j_{1},j_{2})$=\{\textbf{g}:$g_{1,j_{1}}\leq g_{1}\leq h{1,j_{1}+1},g_{2,j_{2}}\leq g_{2}\leq g{2,j_{2}+1}$\}
Also we define 
 $\mathcal{S}(i_{1},i_{2},j_{1},j_{2})$ = $\mathcal{H} \bigcup \mathcal{G}$
 \\
We say that a channel is in state $s_{i_{1},i_{2},j_{1},j_{2}} $if  (\textbf{h},\textbf{g})$ \in \mathcal{S}(i_{1},i_{2},j_{1},j_{2})$
We define quantized power control policy as \\
\begin{equation}
\label{new_power}
P_{k}(h) = \displaystyle\inf_{\textbf{h}\in\mathcal{H}}P_{k}(\textbf{h})
\end{equation}

It can be easily shown that (\ref{p_cons_opt}) implies that (\ref{new_power}) also satisfies power constraint.
Now for a particular quantized state, we have a Gaussian MAC-WT whose rate region is characterized in \cite{yener2006}. In particular, the sum-rate is bounded as:\\
\begin{equation}
R_{1}+R_{2}\leq r_{sum} = \frac{1}{2}\left[log\left(\frac{1+h_{1,i_{1}}P_{1}(h)+h_{2,i_{2}}P(h)}{1+g_{1,j_{1}}P_{1}(h)+g_{2,j_{2}}P_{2}(h)}\right)\right]^{+}
\end{equation}
Now the number of times the channel is in state $s_{i_{1},i_{2},j_{1},j_{2}} $ 
\\
\begin{equation}
N_s = nPr\{(\textbf{h},\textbf{g}) \in \mathcal{S}(i_{1},i_{2},j_{1},j_{2})\}\equiv nPr\{s_{i_{1},i_{2},j_{1},j_{2}}\}
\end{equation}
Now let R$_{sum}$ = R$_{1}$ + R$_{2}$, the following rate is achievable\\
\begin{equation}
\displaystyle\lim_{n\rightarrow\infty} R_{sum} = \displaystyle\lim_{n\rightarrow\infty}\sum\limits_{i_{1}=0}^{M_{1}^{h}} \sum\limits_{i_{2}=0}^{M_{2}^{h}} \sum\limits_{j_{1}=0}^{M_{1}^{g}} \sum\limits_{j_{2}=0}^{M_{2}^{g}} r_{sum}\frac{N_{s}}{n}
\end{equation}
\begin{equation}
=\sum\limits_{i_{1}=0}^{M_{1}^{h}} \sum\limits_{i_{2}=0}^{M_{2}^{h}} \sum\limits_{j_{1}=0}^{M_{1}^{g}} \sum\limits_{j_{2}=0}^{M_{2}^{g}} r_{sum}Pr\{s_{i_{1},i_{2},j_{1},j_{2}}\}
\end{equation}

Due lack of space we omit some steps, one can easily prove also that average probabililty of error also vanishes i.e\\
\begin{equation}
\displaystyle\lim_{n\rightarrow\infty}\bar{P_{e}}\leq\sum\limits_{i_{1}=0}^{M_{1}^{h}} \sum\limits_{i_{2}=0}^{M_{2}^{h}} \sum\limits_{j_{1}=0}^{M_{1}^{g}} \sum\limits_{j_{2}=0}^{M_{2}^{g}} \bar{P_{e}}(i_{1},i_{2},j_{1},j_{2})Pr\{s_{i_{1},i_{2},j_{1},j_{2}}\} = 0
\end{equation}

Now one can show that the optimal power control policy given in theorem 1 achives this sum-rate R$_{sum}^{opt}$ given by\\
R$_{sum}^{opt}$=\\
\begin{equation}
\int_{\tau_{1}}^{\infty}\int_{\tau_{2}}^{\infty}\int_{0}^{\infty}\int_{0}^{\infty} \left[log\left(\frac{1+h_{1}P_{1} + h_{2}P_{2}}{1+g_{1}P_{1} + g_{2}P_{2}}\right)\right]^{+}\phi(\textbf{H})d\textbf{H}
\end{equation}
For that we need to show that for a given ~$\epsilon>0$~$\exists M_{1}^{h},M_{2}^{h},M_{1}^{g},M_{2}^{g}$ s.t\\
\begin{equation}
\sum\limits_{i_{1}=0}^{M_{1}^{h}} \sum\limits_{i_{2}=0}^{M_{2}^{h}} \sum\limits_{j_{1}=0}^{M_{1}^{g}} \sum\limits_{j_{2}=0}^{M_{2}^{g}} r_{sum}Pr\{s_{i_{1},i_{2},j_{1},j_{2}}\} \geq R_{sum}^{opt} - \epsilon
\end{equation}

One can easily show the R$_{sum}^{opt}$ is finite. Then by invoking Dominated convergence theorem along the same lines as done in \cite{gopala2008} the achievability of sum-rate with the optimal power control policy is proved.

\subsection{Proof of theorem 2}
The proof follows along the same lines as in theorem 1 since from \cite{yener2007} we know that when full CSI of eve's channel is known, the optimal power control policy is that one user can either transmit or jam but not both at a time. Hence we can use same arguments as in proof of theorem 1 to prove the achievability of secrecy sum-rate, which we omit here due to lack of space.

\section{Optimal Power Control}
\subsection{Without Cooperative Jamming}
We provide the optimal power control policy without cooperative jamming for Rayleigh fading. Similarly one can obtain the powers for other distributions. Let $f_{1}$ and $f_{2}$ denote the densities of $h$ and $g$ respectively.
For Rayleigh fading case, averaging over all fading realizations of eve's channel, i.e.,  $g_{1}$ and $g_{2}$, which give positive secrecy sum-rate, we get from Theorem 3.1\\ \\
$R$ =
\begin{equation}
\label{optimal_ncj}
\int\limits_{h_{1}}\int\limits_{h_{2}}\left[log\left(\phi_{P_{1},P_{2}}^{h}\right)-\frac{1}{\xi_{P_{1},P_{2}}}\left\lbrace P_{1}\gmga\theta_{P_{1}} - P_{2}\gmgb\theta_{P_{2}}\right\rbrace \right]f_{1}(h)dh
\end{equation}
where\\
$\phi_{P_{1},P_{2}}^{h}$ is as defined in (\ref{notation}) and\\
\begin{equation}
\xi_{a,b} = a\gmga-b\gmgb,
\end{equation}
\begin{equation}
\theta_{P_{1}}=e^{\frac{1}{P_{1}\gmga}}\left[Ei\left(\frac{1}{P_{1}\gmga}\right) - Ei\left(\frac{1}{P_{1}\gmga} + \frac{h_{1}P_{1}+h_{2}P_{2}}{P_{1}\gmga}\right)\right],
\end{equation}
\\
\begin{equation}
\theta_{P_{2}}=e^{\frac{1}{P_{2}\gmgb}}\left[Ei\left(\frac{1}{P_{2}\gmgb}\right) - Ei\left(\frac{1}{P_{2}\gmgb} + \frac{h_{1}P_{1}+h_{2}P_{2}}{P_{2}\gmgb}\right)\right],
\end{equation}
and\\
\begin{equation}
Ei(x)=\int_{x}^{\infty} \frac{e^{-t}}{t}dt.
\end{equation}

After writing the Lagrangian and invoking KKT (Karush-–Kuhn-–Tucker) conditions (which are only necessary here as the objective function need not be concave \cite{boyd}), we get 

\setlength{\arraycolsep}{0.0em}
\begin{eqnarray}
\label{KKT_1}
\frac{h_{1}}{\phi_{P_{1},P_{2}}^{h}} 
&&+ \frac{1}{\xi_{P_{1},P_{2}}}\left\lbrace\frac{\theta_{P_{1}}}{P_{1}}-\gmga+\frac{\alpha_{1}}{h_{2}\phi_{P_{1},P_{2}}^{h}}+\frac{P_{2}\left(\alpha_{1}+\alpha_{2}\right)}{\phi_{P_{1},P_{2}}^{h}}\right\rbrace \nonumber\\
&&+ \frac{P_{2}\gmga\gmgb}{\xi_{P_{1},P_{2}}^{2}}\left(\theta_{P_{1}}-\theta_{P_{2}}\right) - \lambda_{1} = 0
\end{eqnarray}
\setlength{\arraycolsep}{0pt}

\setlength{\arraycolsep}{0.0em}
\begin{eqnarray}
\label{KKT_2}
\frac{h_{2}}{\phi_{P_{1},P_{2}}^{h}}
&&-\frac{1}{\xi_{P_{1},P_{2}}}\left\lbrace\frac{\theta_{P_{2}}}{P_{2}}-\gmgb+\frac{\alpha_{2}}{h_{1}\phi_{P_{1},P_{2}}^{h}}+\frac{P_{1}\left(\alpha_{1}+\alpha_{2}\right)}{\phi_{P_{1},P_{2}}^{h}}\right\rbrace \nonumber\\
&&- \frac{P_{1}\gmga\gmgb}{\xi_{P_{1},P_{2}}^{2}}\left(\theta_{P_{1}}-\theta_{P_{2}}\right) - \lambda_{2}=0
\end{eqnarray}
\setlength{\arraycolsep}{0pt}
where $\lambda_{1}$ and $\lambda_{2}$ are the Lagrangian multipliers and\\
\begin{equation}
\alpha_{1}=h_{2}\gmga e^{-\left(\frac{h_{1}P_{1}+h_{2}P_{2}}{P_{1}\gmga}\right)},
\end{equation}
\begin{equation}
\alpha_{2}=h_{1}\gmgb e^{-\left(\frac{h_{1}P_{1}+h_{2}P_{2}}{P_{2}\gmgb}\right)}.
\end{equation}
\\

We solve this set of equations numerically for optimum power policy:\\
\begin{enumerate}
\item If we find positive solutions for $P_{1}$ and $P_{2}$ from (\ref{KKT_1}) and (\ref{KKT_2}), both should be transmitting with their respective powers.
\item If we do not find positive solutions for both and $h_1 > h_2$, we solve (\ref{KKT_1}) for $P_{1}$ with $P_{2} = 0$.
\item If we do not find positive solutions for both and $h_1 < h_2$, we solve (\ref{KKT_2}) for $P_{2}$ with $P_{1} = 0$.
\end{enumerate}
\subsection{With Cooperative Jamming}
A user can either transmit or jam. We have different expressions of secrecy sum-rate based on whether the users are transmitting or jamming. Averaging (\ref{rs_coop}) over all the fading realizations $(g_{1}, g_{2})$ if there is a positive solution $P_{1}$ and $P_{2}$ from (\ref{KKT_1}) and (\ref{KKT_2}), both users will transmit, and the secrecy sum-rate is given in (\ref{optimal_ncj}).

When there is no solution of (\ref{KKT_1}) and (\ref{KKT_2}) such that $P_{1}>0$ and $P_{2}>0$, and the channel of user 1 is better than that of user 2, the secrecy sum-rate is\\
\begin{equation}
\label{optimal_cj_1}
\int\limits_{h_{1}}\int\limits_{h_{2}}\left[log\frac{\phi_{P_{1},Q_{2}}^{h}}{\phi_{0,Q_{2}}^{h}}-\frac{1}{\xi_{P_{1},Q_{2}}}\left\lbrace P_{1}\gmga\left(\beta_{P_{1}} - \beta_{Q_{2}}\right)\right\rbrace \right]f_{1}(h)dh.
\end{equation}
Similarly when the channel of user 2 is better than user 1, the secrecy sum-rate is\\
\begin{equation}
\label{optimal_cj_2}
\int\limits_{h_{1}}\int\limits_{h_{2}}\left[log\frac{\phi_{Q_{1},P_{2}}^{h}}{\phi_{Q_{1},0}^{h}}-\frac{1}{\xi_{Q_{1},P_{2}}}\left\lbrace P_{2}\gmgb\left(\beta_{Q_{1}} - \beta_{P_{2}}\right)\right\rbrace \right]f_{1}(h)dh.
\end{equation}
where
\\
\begin{math}
\beta_{P_{1}}=e^{\frac{1}{P_{1}\gmga}}\left[Ei\left(\frac{1}{P_{1}\gmga}\right) - Ei\left(\frac{1}{P_{1}\gmga} + \frac{h_{1}}{\gmga(1+h_{2}Q_{2})}\right)\right],
\end{math}
\begin{math}
\beta_{P_{2}}=e^{\frac{1}{P_{2}\gmgb}}\left[Ei\left(\frac{1}{P_{2}\gmgb}\right) - Ei\left(\frac{1}{P_{2}\gmgb} + \frac{h_{2}}{\gmgb(1+h_{1}Q_{1})}\right)\right],
\end{math}
\begin{math}
\beta_{Q_{1}}=e^{\frac{1}{Q_{1}\gmga}}\left[Ei\left(\frac{1}{Q_{1}\gmga}\right) - Ei\left(\frac{1}{Q_{1}\gmga} + \frac{h_{2}}{\gmgb(1+h_{1}Q_{1})}\right)\right],
\end{math}
\begin{math}
\beta_{Q_{2}}~~~=~~~e^{\frac{1}{Q_{2}\gmgb}}\left[Ei\left(\frac{1}{Q_{2}\gmgb}\right) - Ei\left(\frac{1}{Q_{2}\gmgb} + \frac{h_{1}}{\gmgb(1+h_{2}Q_{2})}\right)\right].
\end{math}
\\

Now the problem is to maximize the above objective functions appropriately for each case. This function may not be concave. Hence, KKT conditions are necessary but not sufficient. We solve the equations obtained via KKT numerically to obtain power policy.

\end{document}